\documentstyle[prl,aps,epsfig,multicol]{revtex}
\begin{document}
\def\be{\begin{equation}}
\def\ee{\end{equation}}
\def\bearr{\begin{eqnarray}}
\def\eearr{\end{eqnarray}}

\draft
\title{ Cooperative Ring Exchange and\\
Quantum Melting of Vortex Lattices in Atomic Bose-Einstein Condensates}
\author{Tarun Kanti Ghosh$^{1,2}$ and G. Baskaran$^{1}$}
\address
{$^1$The Institute of Mathematical Sciences, C. I. T. Campus, 
Chennai 600 113, India.
\\$^2$ The Abdus Salam International Centre for Theoretical
Physics, 34100 Trieste, Italy.}

\date{\today}
\maketitle

\begin{abstract}
Cooperative ring-exchange is suggested as a mechanism of quantum melting 
of vortex lattices in a rapidly-rotating quasi two dimensional atomic 
Bose-Einstein condensate (BEC). Using an approach pioneered by 
Kivelson {\em et al}.  [Phys. Rev. Lett. {\bf 56}, 873 (1986)] 
for the fractional quantized Hall effect, we calculate the condition for 
quantum melting instability by considering large-correlated ring exchanges 
in a two-dimensional Wigner crystal of vortices in a strong `pseudomagnetic 
field' generated by the background superfluid Bose particles. BEC may be 
profitably used to address issues of quantum melting of a pristine Wigner 
solid devoid of complications of real solids.

\end{abstract}

\begin{multicols}{2}[]
\section{Introduction}
The creation and observation of the triangular vortex lattices in a 
rapidly-rotating atomic Bose-Einstein condensate (BEC) \cite{mad,abo,eng} 
has opened a new direction for the study of quantum vortex matter.
Theoretical predictions\cite{gunn}
for the existence of fractional quantum Hall like states at even higher
rotational speeds in quasi two dimensional atomic BEC has given a further 
impetus to this fascinating field. The quantum melting of an ordered 
vortex lattice to
an exotic quantum fluid of atoms at very low temperatures is a quantum 
phase transition, where one would like to understand the mechanism of 
melting and nature of phase transition.     

Melting of classical solids with short range inter atomic potential 
in 2D is a well studied subject, where topological defects play a 
fundamental role. In the presence of long range interaction, such as 
one component coulomb plasma in 2D, melting is dominated by ring 
exchanges \cite{choq} rather than topological defects. From this point 
of view the logarithmic repulsion among the imposed vortices in a 
rotating BEC provides an opportunity to study quantum melting of a 
`pristine' Wigner solid with long range forces, that is free from
the complications of solid state systems.

In this letter we write down an effective Hamiltonian for the vortex
degrees of freedom, motivated by an analogy \cite{thouless,don} between 
the Magnus force acting on a vortex moving on a two dimensional neutral superfluid
fluid and the Lorenz force acting on a charged particle in a magnetic field.  
We develop a theoretical approach, borrowing heavily from pioneering ideas 
of Kivelson, Kallin, Arovas and Schrieffer (KKAS) \cite{kiv}, developed in 
the context of fractional quantized Hall effect (FQHE), and suggest a 
cooperative ring exchange (CRE) mechanism for quantum melting of vortex 
lattices in quasi 2D atomic BEC and indicate a possible direction for a 
microscopic understanding of the quantum liquid of molten vortices.   

In contrast to many recent theoretical works on atomic BEC which exploits 
an analogy between the Hamiltonian of a rotating neutral boson atoms and 
charge particle in an external magnetic field in two dimensions, our
work uses the vortex (collective) coordinates directly and provides another 
microscopic approach to understand quantum melting and the quantum Hall-like 
state that may be formed in these atomic system. Existing theoretical works 
focus on exact diagonalization \cite{cooper} of small number of atoms to get
some idea about quantum melting and the possible quantum Hall like 
melted states. 
A recent interesting work \cite{mac} that studies melting
of vortex lattices in a rapidly-rotating 2D BEC, also shows
that BEC is destroyed by the vortex lattices.  

Experimentally, at the present moment it is a challenging task to produce 
vortex liquid state in a rapidly-rotating atomic BEC, in contrast to the 
formation of an incompressible liquid state of electrons in a high magnetic 
field at higher filling fractions. With the rapid advances in the field of 
laser cooled atomic gases one can anticipate to get `snapshots' of the 
melted configurations of the vortex lattices, where CRE should leave its 
unique signatures as we mention at the end.  

In the cooperative ring exchange approach to FQHE, KKAS view the Laughlin 
quantum Hall state as a Wigner solid of electrons in 2D in strong magnetic 
field, that has been quantum melted by cooperative ring exchange 
processes. Briefly, ring exchange, as the name suggests, is a cooperative
shift of a ring of contiguous particles in an ordered lattice (figure 1)
resulting in a cyclic permutation within the ring. While the
amplitude for a quantum tunneling event of a specific ring of size $L$ 
sites is exponentially small $\sim \alpha^L $ (with $\alpha $, the single 
particle tunneling amplitude being $<1$), the number of 
rings of size $L$ is exponentially large $\sim e^{bL}$. Thus the total 
amplitude $\sim \alpha^L e^{bL}$ may exponentially diverge, 
if $ - \ln \alpha < b $, leading to a 
proliferation of ring exchanges and a consequent quantum melting. 

This melting depends on the electronic filling fraction, 
the ratio between the density of conduction electrons and the density of 
flux quanta. For very low filling fraction, electrons are expected to form 
a Wigner crystal. At higher filling fraction, electrons forms an 
incompressible liquid state and  exhibits quantized Hall effect. 
Similarly, we  could also expect the quantum melting of the vortex 
lattices depends on the vortex filling factor, the ratio of the total 
number of vortex to the total number of boson.  

\section{ Hamiltonian of the vortices in a rotating quasi-2D BEC}
We consider a large number of vortices in a rapidly rotating quasi-$2D$ 
BEC; the condition for the condensate in a quasi-2D trap is  
$ \mu = \rho_0 g_2 < \hbar  \omega_z $ and the atoms in the condensate should be in  
the atomic lowest Landau level generated by the fast external rotation 
is $ \rho_0 g_2 < 2 \hbar \Omega $. 
Here, $ \rho_0 $ is the boson density,
$ g_2 = 2 \sqrt{2\pi} \hbar \omega_z a_z a $ is the effective 
interaction strength in quasi-2D Bose system 
\cite{tin}, where $ a_z = \sqrt{\frac{\hbar}{m_o\omega_z}}$ is 
harmonic oscillator length along the $z$-direction with $ \omega_z$ 
is trap frequency in the axial ($z$) direction and $ \Omega $ is 
the trap rotational frequency. Also, $ a$ is the $s$-wave atomic scattering length.
A vortex in a fluid is an excitation in which each fluid
particle is given an angular momentum $m$ relative to 
the vortex center. Here, we treat a vortex as a point 
particle moving under the influence of the Magnus force.
The Magnus force is an effective interaction between superfluid particles 
and vortices in relative motion\cite{thouless,don}.
The force acting on a single vortex\cite{thouless} is then
\begin{equation}
{\bf F} = {\bf v} \times \hat{z} (2\pi \hbar \rho_0).
\end{equation}
Here, $ {\bf v}$ is the vortex velocity relative to the superfluid
particles and $ \rho_0 $ is the superfluid particle density.
The Magnus force is equivalent to the Lorentz force acting
on a charge particle ($ e $) in a magnetic field. Hence, 
$ e B_{\rm eff} = 2 \pi \hbar \rho_0 $ is the pseudo magnetic field.

The interaction potential between two vortices separated 
by a distance $r$ is 
\be
V(r) = - \frac{2 \pi \hbar^2 \rho_0}{m_0} \ln \left (\frac{r}{\xi} \right ),
\ee
where $ \xi \sim \sqrt{\frac{a_z}{\rho_0 a}} $ is the 
coherence length of the vortex core and $ m_0 $ is 
the mass of a superfluid particle. The above potential is 
valid only when the distances between two vortices 
is greater than the coherence length. 
Notice that the interaction strength between two 
vortices depends on the superfluid density as well 
as the $s$-wave scattering length $a$.  

The Hamiltonian of a rotating BEC containing vortices can be written in 
terms of center of vortices (collective coordinate) as \cite{thouless}  
\be
H_v = \sum_{i=1}^{N_v} \frac{ ({\bf p}_i - \pi \hbar \rho_0 \hat {z} \times 
{\bf r}_i )^2}{2 m_v}
- \frac{2 \pi \hbar^2 \rho_0}{m_0} \sum_{i<j} \ln \left [ \frac{ |{\bf r}_i - {\bf r}_j |}{\xi} \right ],
\ee
where $ N_v $ is the total number of vortices.
The effective vortex mass  $ m_v = \pi \rho_0 \xi^2 $ can be in principle derived
from a microscopic approach \cite{thouless}. Since the coherence length
is very small, the vortex mass also becomes small.  
This Hamiltonian is similar to that of a charged
particles moving under the influence of the Lorentz 
force by a magnetic field $ B_{\rm eff}$. The pseudo 
vector potential due to the Magnus force is
${\bf A}_{\rm eff} = - \frac{1}{2} {\bf r} \times {\bf B}_{\rm eff}$.
For $ N_v $ number of vortices in an area $ A $, 
one gets the vortex filling factor,
$\nu_v = \frac{N_v}{A} \frac{ h}{e B_{\rm eff}} = \frac{N_v}{N}$.
Notice that the vortex filling factor $ \nu_v$ is just 
inverse of the bosonic filing factor $\nu_b = \frac{N}{N_v}$.
For large $ N_v $ the vortex density is approximately 
uniform and $ N_v = \frac{2 m \omega A}{h} $.
$ N $ is the number of the superfluid particles. 
The effective magnetic length is
$ l_0 = \sqrt{\frac{\hbar }{ e B_{\rm eff}}} =
\frac{1}{\sqrt{2 \pi \rho_0}} $.
The pseudomagnetic field generated by the background superfluid
particles leads to the quantization of the cyclotron motion and
producing Landau levels of the vortices. 
The eigen spectrum of the single vortex Hamiltonian are 
uniformly spaced with energy gap $ \hbar \omega_{\rm eff} $, 
where $ \omega_{\rm eff} = \frac{2\pi \hbar \rho_0}{ m_v} $ is 
the effective cyclotron frequency.  The limit of 
$ m_v \rightarrow 0 $ and/or large superfluid density ($ \rho_0$)
is equivalent to the vortices are in the lowest Landau level (LLL). 
We can project the Hamiltonian onto the LLL and the 
corresponding normalized wave functions are degenerate eigenfunctions 
of the angular momentum $ m $ is, 
\be
\psi(z) = \frac{1}{\sqrt{(2l_0^2)^{m+1} \pi m! }} z^m e^{-\frac{|z|^2}{4 
l_{0}^2}} , \hspace{0.5 cm}  m = 0, 1, 2,....
\ee
where $ z = x+i y $ and $(x,y)$ are the position coordinate of a vortex. 
When the vortices are confined to 
the lowest Landau level,(i.e. the cyclotron degrees of freedom are
confined to the LLL), the kinetic degrees of freedom
of the vortices are frozen, since the spacing 
between Landau levels, $ \hbar \omega_{\rm eff}$,
is large compared with all other energies in 
the problem.
The remaining degrees of freedom are the vortex guiding-center coordinates,
$ {\bf R} = \frac{{\bf r}}{2} + \frac{l_0^2}{\hbar} ({\bf p } \times \hat {z})$.
The guiding center coordinate ${\bf R} $
specify the center of a Gaussian-localized probability amplitude
of width $l_0$. These coordinates have no kinetic energy. 
Hence the vortices in the LLL will remain localized  about a given guiding center 
coordinate $ {\bf R}$ indefinitely. 

\section{Coherent state path integration}
In this section, we would like to review the coherent state path integral
formalism. For detailed derivations, please consult the references \cite{sch,gir}.
In symmetric gauge, the wave function of a vortex 
in the LLL with guiding-center position $ {\bf R}$ is
\be
 \phi_{{\bf R}} ({\bf r} ) = \frac{1}{\sqrt{2\pi l_0^2}}
 exp \left [- \frac{| {\bf r} - {\bf R} |^2}{4 l_0^2} 
 + \frac{i ( {\bf r} \times {\bf R} ).\hat z}{2 l_0^2} \right ].
\ee
It has the same form as a coherent state in a 
two-dimensional phase space \cite{sch}.
Here, the state label $ {\bf R}$ is a continuous variable. 
The coherent state overlap is given by,
\be
<{\bf R}_1 | {\bf R}_2 > = exp \left [- \frac{ |{\bf R}_1 -
{\bf R}_2 |^2}{4l_0^2}
+ \frac{i ({\bf R}_1 \times {\bf R}_2 ).
\hat z )}{2 l_{0}^2 } \right ].
\ee
This coherent state $ |{\bf R} > $ forms a
nonorthogonal, overcomplete basis.
Nevertheless, the projection operator $ P $ onto
the LLL is given by,
\be
P = \int \frac{ d^2 R }{2\pi l_{0}^2} | {\bf R} >< {\bf R} |
\ee
which is unity within the LLL since
$ < {\bf R}_1 | P | {\bf R}_2 > = <{\bf R}_1 | {\bf R}_2> $.

We use the coherent state path integral \cite{sch,gir}
expression for the partition function to calculate the
tunneling coefficient of a vortex.  The partition
function for $ 2D $ interacting vortices in a pseudomagnetic
field due to the Magnus force is
\be
Z (\nu_v ) = Tr e^{- \beta H_v }.
\ee
Here, we discuss the main features of this formalism
for a single vortex in the LLL in the complex plane.
This can be generalized for many vortex system very
easily. The coherent state in the complex plane is
\be
| R> = \frac{1}{\sqrt{2 \pi l_{0}^2}}
exp \left [\frac{z R^* - z^* R}{4 l_{0}^2} \right ]
exp \left [-\frac{| z-R|^{2}}{2l_{0}^2} \right ],
\ee
where $ R = X+iY $ is the guiding center coordinate
of a vortex in the complex plane and the asterisk denotes
the complex conjugation.
The coherent state overlap in the complex plane is
\bearr \label{overlap}
<R_j|R_k>&=& exp \left [(\frac{R_j R_{k}^* - R_{j}^* R}
{4 l_{0}^2}) - \frac{| R_j-R_k|^{2}}{2l_{0}^2} \right ]
\\ &=& exp \left [-\frac{
[R_j^*(R_j - R_k) - (R_{j}^* - R_{k}^*)R_k}{4 l_{0}^2} \right ].
\eearr
Now the path integral representation of the partition
function $ Z (\nu_v) $ can be obtained in the usual way.
First, we split the inverse temperature
$ \beta $ into a large number of equal intervals
$ \epsilon  = \beta /n $, i.e., $ e^{- \beta V } $ is
written as  $ [e^{- \epsilon V}]^n $, and then insert
the projection operator $ P $ at each infinitesimally small
interval. Then,
\be
< R_f | e^{- \beta V } |  R_i> =
 \int \prod_{k=1}^{n} \frac{ d^2 R_k}{(2\pi l_{0}^2)^n} \prod_{j=0}^{n}
<  R_{j+1} |  e^{- \epsilon V } |  R_{j}>,
\ee
where $ R_0 =  R_i, R_{n+1} = R_f$.
In general, the Hamiltonian can be written
\be
V_{j,k} = V(R_j^{*}, R_k) = \frac{<R_j|V|R_k>}{<R_j|R_k>}.
\ee
The matrix element can be written as,
\bearr
< R_f | e^{- \beta H } | R_i> & = &  \nonumber
\int \prod_{k=1}^{n} \frac{ d^2 R_k}{(2\pi
l_{0}^2)^n} \prod_{j=0}^{n} < R_{j+1} | R_{j}> \\ & \times &
 \left [ 1 - i \epsilon V ( R_{j+1}^*, R_{j}) + O(\epsilon^2) \right ]
\eearr
We are neglecting terms of $ O (\epsilon^2) $ and higher
order terms by standard procedure.
Using Eq.(\ref{overlap}), and
\be
\frac{dR_j}{dt} = \frac{ R_{j+1} - R_j}{\epsilon}
\ee
we obtain
\be
Z(\nu_v)  =   \nonumber \int \prod_{k=1}^{n}
\frac{ d^2  R_k}{(2\pi l_{0}^2)^n}
exp \left [ i \epsilon \sum_{j =0}^n \frac{i}{4 l_{0}^2}
S_j \right ],
\ee
where 
\be
S_j =\left ( R_j^{*} \frac{dR_j}{dt} -
\frac{dR_j^{*}}{dt} R_{j+1} \right ) -  V (R_j^{*}, R_{j+1}).
\ee
The above path integral can be written as
\be
Z = \int D[{\bf R}] e^{i S[{\bf R}]},
\ee
where 
\be
S = \int_{0}^{\beta} dt \left [\frac{1}{4 l_{0}^2}
\left (R^{*} \frac{dR}{dt} -
\frac{dR^{*}}{dt} R \right ) +  V (R^{*}, R ) \right].
\ee
This action is linear in time derivatives and hence
discontinuous paths have finite action.
It implies that the coherent state path integral is
dominated by discontinuous paths and the
limits is ill defined. Despite these difficulties,
the continuum version of the path integral
can be used to develop a saddle-point approximation
for the partition function \cite{kiv}.
We are interested in the semiclassical limit when $ V(R) $
is a slowly varying function of its argument over
the length scale $ l_0 $ and we can use the saddle
point approximation to evaluate the path integral.

The single vortex path integral can be generalized
to many vortex path integral directly. The action for many vortex is
\be
S ({\bf R}) = \int_{0}^{\beta} dt
\left [ \frac{1}{2} \sum_{j = 1}^{N_v}
 \left ( \dot{{\bf R}}_j \times {\bf R}_j \right ). \hat z
+ \sum_{j<k} V \left ({\bf R}_j - {\bf R}_k \right ) \right ],
\ee
where
\be \label{pot}
 V (R) = \langle \phi_{{\bf R}}({\bf r})| V({\bf r}) |
\phi_{{\bf R}} ({\bf r}) \rangle 
\ee
is two-body interaction potential in coherent states representation. 

In the saddle point-approximation, the classical path is 
obtained by minimizing the action, 
$ \frac{\delta S}{\delta R_j(\tau)}|_{R=R_c} = 0 $. 
The classical paths satisfy the following equations of motion
\be\label{dy}
\dot {{\bf R}}_j = \frac{l_0^2}{\hbar} (\nabla_j V_j) \times \hat z,
~{\rm where}~ V_j = \sum_{k \neq j } V({\bf R}_j - {\bf R}_k).
\ee
 
The path integral can be expressed as a sum over saddle point 
contribution in which the contribution of 
paths in the neighborhood of each classical path
is evaluated by expanding the action
to quadratic order in $ R - R_c $.
The partition function $ Z $ is calculated 
within the semiclassical approximation. The partition 
function can be expressed as a sum over classical paths, assuming
the vortices to be bosons, 
$ Z = \sum_{c} D[R_c] e ^{-S[R_c]} $,
where $ D[R_c] $ is the fluctuation determinant and $ S[R_c] $ is the 
action evaluated along the classical path. There are interesting 
issues\cite{hal}
about the statistics of vortices in a compressible superfluid such as ours,
and in the present paper we assume them to be bosons.
The partition function $ Z $ can be written by considering only
the leading order contribution as \cite{kiv}
$ Z = Z_0 \sum_c \tilde{D}[R_c] e^{-S_0[R_c]}$,
where $ {S} =  S_0 + \tilde S $.
Let us consider the contribution of a single large 
exchange ring to $ Z $. The real part of the action 
would be $ \alpha_0 L $ (see the sec. V), where $ L $ is the number 
of vortices in the ring and $ \alpha_0 $ is independent 
of path. The fluctuation determinant \cite{kiv} is
$ \tilde{D}[R_c] \sim exp[-\delta \alpha L + O (\ln L)]$.
Here, $ \delta \alpha $ is a real constant which renormalize $ \alpha_0$.
The imaginary part, the phase change as a cooperative
motion along a ring is 
$ \theta = \oint e {\bf A}_{\rm eff} . d{\bf l} = 2 \pi N $, 
where $ N $ is the number of the superfluid particles 
enclosed by the ring \cite{hal}. This is the analog of 
the Bohm-Aharanov phase factor for a charged particle moving
in a magnetic field. So the partition function becomes,
$ Z \sim exp[- \alpha L \pm i 2\pi N ]$, 
where $ \alpha = \alpha_0 + \delta \alpha $.

\section{Cooperative ring exchange mechanism}
How does the vortex lattice melt? To understand the melting of the vortex
lattices, the Lindemann criteria can not be used here since it is
used in the melting of classical solids.  
The vortices are not executing almost independent 
thermal motions as in a classical solid. The dynamics
of the present problem is governed by a Hamiltonian 
with only first-order time derivatives, which give 
rise to its own peculiar properties. If we
consider a rigid Wigner solid and allow one ring  
of vortices to tunnel coherently they see a periodic 
potential with the periodicity of the lattice (Fig 1).
If we observe the coherent motion of one chain 
over a long time compared to the tunneling time ($ \tau_0 $), the potential 
that it sees will not be periodic.
The physically important rings being one-dimensional 
and long, this can result in the destruction of 
the long-range order along the chain rather easily.
This in turn will feed back and affect the rest of 
the neighborhood, resulting possibly in a molten state. 
This will also result in the path of the wave packets
of vortices being displaced away from the edges of the 
triangle of the lattice. This means that the self-consistent
potential seen by a vortex no longer has a component 
which has long-range order.

\begin{figure}[h]
\epsfxsize 8cm
\centerline {\epsfbox{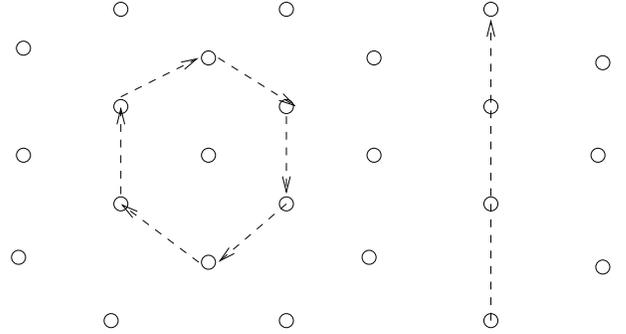}}
\vspace{0.5 cm}
\caption{ A schematic  diagram of cooperative ring exchange events on a ring
and line. The vortices are indicated by the small circles.}
\end{figure}

\section{calculation of the tunneling coefficient}

The numerical value of the tunneling coefficient 
$ \alpha (\nu_v ) $ determines whether the vortices 
form the liquid state or Wigner crystal. To estimate 
this tunneling coefficient we consider the following 
simple exchange path which is shown in figure 1. Consider 
the path in which one row of vortices exchanges one 
step in the $ X $ direction in the background of 
the static potential of all other vortices,
$  X_i(\beta) = X_i(0) + d $ and $ Y_i(\beta) = Y_i(0) $,
where $ d = \sqrt{\frac{4 \pi }{\sqrt{3} \nu_v}} l_0 $ is
the lattice constant of the Wigner crystal for a given 
density $ \nu_v $. There is no net phase changes since 
this straight path does not enclose any area.
We are imposing the periodic boundary conditions 
in the $ X $ direction, $ X_i(\tau) = X_{i+ L}(\tau) $.

For $ |Y_j| \ll  d $, the two-body interaction potential in 
the coherent state representation (given in the eq. \ref{pot}) can be 
approximated by 

\bearr
\frac{V}{\hbar \Omega} & = & \frac{4 \pi }{\sqrt{3} \nu_v } \{ \sum_{j = 1}^{L} \left [ 
\frac{Q_y}{2} Y_j^2 + \frac{Q_x}{(2\pi)^2} (1-\cos(2\pi X_j)) \right ]
\\ \nonumber & + & \frac{1}{2} \sum_{j>k}[K_x(j-k)(X_j - X_k)^2 + K_y(j-k) (Y_j - Y_k)^2] \},
\eearr
where $ X_j'$s  and $ Y_j'$s are in units of the lattice 
constant $ d $. 
Here, $ K_x(j-k) = \frac{\partial^2V_{jk}}{\partial X_j \partial X_k}|_{R_C} $ and
$ K_y(j-k) = \frac{\partial^2V_{jk}}{\partial Y_j \partial Y_k}|_{R_C} $ 
are evaluated along the classical paths.
The best fit to the actual potential is 
obtained with $ Q_x / Q_y \sim 0.6 $ for $ \nu_v \sim 1/2 $
and $  Q_x $, $ Q_y $ and $ K_x(j) $, $ K_y(j) $ are weakly 
dependent on $ \nu_v $.
The calculation of the fitting parameters $  Q_x $ and $ Q_y $ is given in the
Appendix. Notice that $ Q_x/Q_y  < 1 $ implies
that when one-dimensional chain moves coherently in the 
$ X $ direction, the potential barrier is much less than 
in the $ X $ direction compared to that of the $ Y $ direction.

The dimensionless Euclidean action is given by
\bearr\nonumber
S & = & \frac{4 \pi }{\sqrt{3} \nu_v }
\int_{0}^{\beta} d \tau  [ \sum_{j} [ \frac{i}{2} \dot X_j Y_j + 
\frac{Q_y}{2} Y_j^2 
\\ \nonumber & + & \frac{Q_x}{(2 \pi)^2} (1-\cos(2 \pi X_j)) ]
\\ \nonumber & + & \frac{1}{2} \sum_{j>k} [K_x(j-k)(X_j - X_k)^2 + K_y(j-k) (Y_j - 
Y_k)^2 ]]. 
\eearr
 Since $ S $ is a quadratic form in $ Y_j $, 
the motion in the $ Y $ direction can be integrated 
out exactly. After doing the $ Y_j $ integration, 
we get an effective action $ S_{\rm eff} $ for the 
$ X $ motion, with a quadratic kinetic energy,

\bearr
S_{\rm eff} & = & \frac{ 1 }{\sqrt{3} \pi \nu_v } \int_{0}^{\beta} d \tau [
\sum_{j<k}[ \frac{1}{2} \dot {\phi_j} M(j-k) \dot {\phi_k}
\\ \nonumber & + & \frac{1}{2}
[K_x(j-k)(\phi_j - \phi_k)^2] + \sum_j \frac{Q_x}{(2\pi)^2}(1-\cos(\phi_j))],
\eearr
where $ \tau $ is the imaginary time variable, 
$( M(j-k))^{-1} = \frac{1}{8} [ (\frac{Q_y}{2} + \sum_j K_y(j) 
\delta_{jk})- \sum_{j<k} K_y(j-k)] $ and $ \phi_j = 2 \pi X_j $.
$ S_{\rm eff} $ is the effective action for a one-dimensional sine-Gordan chain. The 
classical path satisfying the boundary conditions are $ \phi_j (0)=0 $ 
and $ \phi_{j}(\beta) = 2 \pi $, corresponds to the simultaneous coherent motion
of all the vortices, i.e. $ \phi_j(\tau) = \phi_{0} (\tau) + 2\pi j $. 
Due to the simultaneous coherent motion of all the vortices, the above effective 
action becomes,
\be
S_{ \rm eff} = \frac{1}{\sqrt{3} \pi \nu_v} \int d \tau \sum_{i}^{L}
 \left [ \frac{\dot {\phi_i}^2}{8 Q_y} + Q_x (1 - \cos(\phi_i)) \right ].
\ee
By using the Euler-Lagrange equation of motion, one can calculate the $ \dot {\phi}$.
Hence, the above effective action along the classical path becomes, $ S[R_c] = \alpha_0 L $,
where $ \alpha_0(\nu_v) = \frac{4}{\sqrt{3}\pi \nu_v} \sqrt{\frac{Q_x}{Q_y}} $.
The $ \alpha_0 $ is independent of $ K's $. 
To evaluate the fluctuation determinant we have to take the continuum limit
of the effective action $ S_{\rm eff}$. To take a continuum limit of the effective action,
$( \phi_j - \phi_k)$ is replaced by $ (j-k) \partial_x \phi $, but $\sum_j j^2 K_x(j)$
is diverging linearly since $ K_x(j) \sim 1/j^2 $. 
This is an infrared divergence and the continuum model must be
constructed by taking the upper cut-off limit carefully. 
Here, we do not calculate the $ \delta \alpha $ which is a non-trivial task.

Kivelson {\em et al}. \cite{kiv} has given an extensive discussion of
how to map Wigner crystal of electrons in a magnetic field into the discrete 
Gaussian model. 
Following the ref. \cite{kiv}, one can map the sum over all classical
paths to a sun over classical spin configurations.
All the contributions of ring exchanges happening in a time interval
$ \tau_0 $ are summed by modeling the change in the action by a 
discrete Gaussian model in an imaginary field \cite{chu},
\be
H_{DG} = \alpha_0(\nu_v) \sum_{<\lambda,\gamma>} (S_{\lambda} - S_{\gamma})^2 
+ i h(\nu_v) \sum_{\lambda} S_{\lambda},
\ee
where $ <\lambda,\gamma> $ denotes a nearest-neighbor pair
on the dual lattice and  $ S_{\lambda} $ is an integer 
variable associated to
every triangle in the lattice. $ S_{\lambda} $ counts the
number of clockwise minus counterclockwise ring exchanges that
surround a plaquette $ \lambda $. The function $ \alpha_0(\nu_v) $
is a measure of the tunneling barrier. The function 
$ h(\nu_v)$ is the phase factor 
which arises as a result of the pseudo magnetic flux enclosed
by the exchange rings.
This model is known to have a phase transition
\cite{chu} at a critical value of $ \alpha = \alpha_c(\nu_v) \sim 1.1 $
\cite{kiv}. For $ \alpha (\nu_v) > \alpha_c (\nu_v) $, the ground state is
a vortex Wigner crystal and for $ \alpha (\nu_v) < \alpha_c (\nu_v)$,
the ground state is a quantum mechanical vortex liquid state.
In our calculation we find that the quantum melting will 
occur at $ \nu_v \sim \frac{1}{2} $. 
The current experiments \cite{mad,abo} with $ \nu_v \ll \frac{1}{2} $ 
are in the regimes of vortex lattice ground state.
So our result is consistent with the experimental results,
but it does not match very well with the other theoretical
results \cite{cooper,mac}.
The numerical calculation is based on the small number of atomic
bosons as well as small number of vortices. In our approach
we assumed a large number of atomic bosons and vortices. We believe
that this discrepancies is related to the system size.
One would say that a large system has been considered in the ref. \cite{mac} and
calculated the melting condition which is comparable to the numerical result \cite{cooper}.
In the ref. \cite{mac}, first they have calculated the root mean square of the 
displacement from the equilibrium position of a vortex in terms of the filling factor $\nu_v$.
Then they have used the Lindemenn criterion and assumed that the melting will occur when the fluctuation
of the vortex position is $ 0.15 d $ to get the melting condition which is
close to the numerical result \cite{cooper}. 
Although the Lindemann criterion gives a reasonable description of the melting of a classical solid,
there is little evidence that it can be applied to the melting of a vortex lattice. The vortices
are intrinsically quantum objects whose equation of motion are quite different from those of atoms
in a harmonic crystal.

\section{summary}
In this paper, we treated the vortices as a new degrees of freedom and  
considered a model Hamiltonian of interacting vortices. Later, we assumed
the vortices are in the lowest Landau level due to the low mass of
the vortices and the high densities of the superfluid Bose particles.
The concept of cooperative ring exchange is introduced
to explain the mechanism of quantum melting of the Wigner crystals.
Finally, we estimated the tunneling coefficient which determines 
the  condition for quantum melting instability of the vortex lattices.
Latest experiments \cite{mad} with $ N \sim 10^5 $,  $ N_v \sim 10 $ 
($ \nu_v \sim 10^{-4} $) and \cite{abo} with $ N \sim 10^{7} $,  $ N_v \sim 100 $ 
($ \nu_v \sim 10^{-5}$) are in the regime in which the ground state is a vortex lattice.
It is a challenge  for experimentalists to produce 
a vortex liquid state in a rotating Bose condensed state.

Our present work, resulting in a discrete Gaussian model (equation 10) 
predicts Laughlin like even denominator bosonic vortex filling 
fraction $\nu_v = \frac{1}{2}$, to emerge on 
quantum melting. We can also determine the asymptotic form of the wave 
functions\cite{dhlee}. Along with a rich phase structure the discrete
Gaussian model also  
determines the nature of the quantum melting transition. To the extent 
the vortex degrees of freedom retain their identity, the results of CRE 
approach may remain valid in the quantum melted region.  This needs to
be investigated further. 

As mentioned earlier CRE processes should leave its finger print as 
specific fluctuation patterns (figure 1) that preempts quantum melting. 
It should be interesting to look for  snapshots of such displaced large
rings in the actual vortex lattice imaging. 


\begin{appendix}
\section{Calculation of the parameters $Q_x$ and $Q_y$}
Here, we describe how to calculate the parameters $ Q_x $ and $ Q_y $.
We consider the simplest possible exchange path, namely
one line of vortices shifting coherently within the Wigner crystal.
When the line $ \cal L $ is displaced, we have
$ {\bf R}_i = {\bf T}_i + {\bf d} \delta_{i \in \cal L}$,
with $ \delta_{i \in \cal L} $ unity if and only if lattice site $ i $
lies on the line in question.
The matrix element of the potential between two vortices in coherent basis
state is
\be
V( {\bf R}) = \langle \phi_{{\bf R}}({\bf r}) | V({\bf r})| \phi_{{\bf R}}({\bf r}).
\rangle 
\ee
Accordingly, the energy of the displaced line configuration relative to that
of the perfect Wigner crystal is
\be
\Delta E = \sum_{\langle i,j \rangle}  \left [ V ( {\bf R}_i - {\bf R}_j) - V ({\bf
T}_i - {\bf T}_j) \right ].
\ee
This sum can be broken up into three terms. The first term includes all
pairs $(i,j)$ in which both sites $ i $ and $ j $ lie off the line. This
contribution  to $ \Delta E $ is zero.  The second term involves
all pairs $(i,j)$ where one of the sites, say $ i $, is on the line
and the other, $ j $, is off the line:
\be
\Delta E_2 = \sum_{i \in \cal L, j \notin \cal L } \left [ V ( {\bf T}_i +{\bf d} -
{\bf T}_j) - V ({\bf T}_i - {\bf T}_j) \right ].
\ee
Clearly the line energy is extensive, hence the energy per tunneling of the vortex
can be written
\be
U({\bf d}) = \Delta E_2 / L =
\sum_{j \notin \cal L  } \left [ V ( {\bf T}_j - {\bf a})
- V ( {\bf T}_j) \right ],
\ee
where we have chosen the origin to lie on the line. The
third and final term is that arising from both $ i $ and $ j $
on the line. Since the tunneling is cooperative, this contribution
to the classical action vanishes.

By allowing one line of vortices to tunnel coherently along the line,
one can fit the change in energy $\Delta E$ into a periodic potential with
the appropriate choice of the parameter $ Q_x $.
On the other hand, by allowing one line of vortices to tunnel coherently
perpendicular to the line, one can fit the change in energy into a
quadratic potential with appropriate choice of the parameter $ Q_y$.

\end{appendix}

\end{multicols}

\begin{references}


\bibitem{mad}
 K. W. Madison, F. Chevy, W. Wohleben, and J. Dalibard,
Phys. Rev. Lett. {\bf 84}, 806 (2000).

\bibitem{abo}
J. R. Abo-Shaeer, C. Raman, J. M. Vogels, and W. Ketterle, Science {\bf 292},
476 (2001).

\bibitem{eng}
P. Engels, I. Coddington, P. C. Haljan, V. Schweikhard, and E. A. Cornell,
 Phys. Rev. Lett. {\bf 90}, 170405 (2003).

\bibitem{gunn}
N. K. Wilkin and J. M. F. Gunn, Phys. Rev. Lett. {\bf 84}, 6 (2000); 
S. Viefers, T. H. Hansson and S. M. Reimann, Phys. Rev. A {\bf 62}, 053604 (2000);
Tin-Lun Ho, Phys. Rev. Lett. {\bf 87}, 060403 (2001).

\bibitem{choq}
Ph. Choquard, and J. Clerouin, Phys. Rev. Lett. {\bf 50}, 2086 (1983).

\bibitem{thouless}
Q. Niu, P. Ao, and D. J. Thouless, Phys. Rev. Lett. {\bf 72}, 1706 (1994).

\bibitem{don}
R. J. Donnelly, {\em Quantized Vortices in Helium II}, (Cambridge, 1991).
 
\bibitem{kiv}
S. Kivelson, C. Kallin, D. P. Arovas, and J. R. Schrieffer, Phys. Rev. Lett.
{\bf 56}, 873 (1986); {\em ibid},
Phys. Rev. B {\bf 36}, 1620 (1987).

\bibitem{cooper}
N. R. Cooper, N. K. Wilkin, J. M. F. Gunn, Phys. Rev. Lett.
{\bf 87}, 120405 (2001).


\bibitem{mac}
Jairo Sinova, C.B. Hanna, A. H. MacDonald, 
Phys. Rev. Lett. {\bf 89}, 030403 (2002).

\bibitem{tin}
Tin-Lun Ho and Michael Ma, J. Low Temp. Phys. {\bf 115}, 61 (1999).

\bibitem{sch}
L. Schulman, {\em Techniques and Applications of Path Integration},  
(Wiley New York, 1981).

\bibitem{gir}
S. M. Girvin and T. Jach, Phys. Rev. B {\bf 29}, 5617 (1984).
 
\bibitem{hal}
R. Y. Chiao, A. Hansen and A. A. Moulthrop, Phys. Rev. Lett. {\bf 54}, 
1339 (1985); F. D. M. Haldane, and Y. S. Wu, Phys. Rev. Lett. {\bf 55}, 2887 (1985).

\bibitem{chu}
S. T. Chui and J. D. Weeks, Phys. Rev. B {\bf 14}, 4978 (1976);
W.Y. Shih and D. Stroud, Phys. Rev. B {\bf 32}, 158 (1985) 

\bibitem{dhlee} Dung-Hai Lee, G. Baskaran and S. A. Kivelson,
Phys. Rev. Lett. {\bf 59}, 2467 (1987).  

\end{references}
\end{document}